\documentclass[prd,11pt,aps,floats,nofootinbib,tightenlines,showpacs]{revtex4-1}
\usepackage{epsfig,graphicx,pstricks}
\usepackage{physics}
\usepackage{psfrag}
\usepackage{color}
\usepackage{amsmath}
\usepackage{mathtools}
\usepackage{amsfonts}
\usepackage{amssymb}
\usepackage{textcomp}
\usepackage{multirow}
\usepackage{subfigure}

\usepackage{epsfig,graphicx,pstricks}
\begin{document}
	\title {State-Dependent Quantum Copying: an adaptive ancillary systems and its limitations }
		\author{Guruprasad Kadam }
		\email{guruprasadkadam18@gmail.com}
		\affiliation{Department of Physics and Materials Science and Engineering,\\
			Jaypee Institute of Information Technology,\\
			 A-10, Sector-62, Noida, UP-201309.}
\date{\today}

\begin{abstract}
In this work, we introduce a novel state-dependent quantum cloning (copying) process by introducing a new class of ancillary system---an {\it{adaptive ancilla}}---modifying the conventional state-dependent quantum  copying process. This state-dependent ancillary system is not pre-engineered to match the quantum state to be cloned, rather it dynamically aligns with the quantum state to be cloned via interaction. However, the space of states that it can clone is restricted by the symmetry principles. This process, while resembling quantum cloning, adheres to the no-cloning theorem due to its state-dependent and non-universal nature. Also, no-cloning theorem does not forbid the possibility that the information required to construct a clone pre-exist in any implicit form but forbids  the construction of a new copy using single universal cloning machine or the existence of a hidden copy. We clarify the distinction between universal copying and conditional copying, and also between state-dependent copying via pre-engineered ancilla and  via {\it{adaptive ancilla}}. We demonstrate that stimulated emission offers a concrete physical realization of state-dependent quantum copying via {\it{adaptive ancilla}}.  We explore how a quantum state, for instance a photon polarization, can be  cloned through light-matter interactions when the ancillary system, such as an excited atom, contain an  implicit structural information about the quantum state in the form of structured set of dynamical response channels.   We reinterpret the excited atomic state  as a realization of an \emph{adaptive ancilla} and cloning of a photon polarization state occurs when the quantum state of an excited atom dynamically aligns with the polarization state of the photon through physical interaction.   We demonstrate that the true limits of cloning arise  solely not from the no-cloning theorem, but from the symmetries imposed on physical systems\textthreequartersemdash constraints which may, in principle, be relaxed or engineered in suitable quantum systems, for instance in Rydberg atoms. 
\end{abstract}

\maketitle

\section{Introduction}

With the recent advances in the development  of quantum technologies, namely quantum computing \cite{aharonov1999quantum,nielsen2010quantum}, quantum communication \cite{gisin2002quantum,pirandola2020advances}, it becomes of utmost important to understand the limits of quantum mechanics as far as manipulating quantum information is concerned. No-cloning theorem plays a central role in this endeavor. It states that it is impossible to construct a universal unitary operator that can clone an arbitrary quantum state. That is, there exists no unitary operator \( \mathcal{U} \) such that:
\[
|\Psi\rangle \otimes |A\rangle \xrightarrow{\mathcal{U}} |\Psi\rangle \otimes |\Psi\rangle
\]
for all \( |\Psi\rangle \), where \( |A\rangle \) is a fixed ancilla independent of the state. 

To quote Asher Peres \cite{peres2003history}, "...these things were well known to those who knew things well". Formal proof of the theorem, however, came much later \cite{wootters1982single,Dieks:1982dj,yuen1986amplification}.  Proof was inspired by a superluminal communication device called FLASH proposed by Nick Herbert which could be used to send  coded messages to distant locations with superluminal speeds \cite{herbert1982flash}. The device was an idealized laser gain tube which would give a macroscopically distinguishable output for a single arbitrary polarized photon input. Thus, the working of the device relies on the cloning of an arbitrary photon polarization state. However, proof of the no-cloning theorem shows that the physical operation of universal cloning is inconsistent with the linearity of quantum mechanics. Hence, such device cannot be an ideal cloning machine.

 No-cloning theorem also inspired other no-go theorems, namely, no-broadcasting theorem\cite{barnum1996noncommuting},  no-deleting theorem \cite{kumar2000impossibility} and no-hiding theorem \cite{braunstein2007quantum}. For an extensive review of quantum cloning, reader may refer  \cite{scarani2005quantum,fan2014quantum}.  

However,  no-cloning theorem does not prohibit cloning under certain conditions:
\begin{itemize}
    \item If the set of states is finite and known (state-dependent cloning),
    \item If the ancilla contain an {\it{implicit}} information about \( |\Psi\rangle \) (not a pre-existing copy),
    \item Or if the cloning is approximate or probabilistic.
\end{itemize}
In this context, various cloning machines have been proposed, namely probabilistic cloning \cite{duan1998probabilistic}, optimal cloning \cite{buvzek1996quantum, gisin1998quantum} etc. In case of probabilistic cloning, ancilla states are pre-enginnered and the cloning works only for finite, known set of states. In case of optimal cloning, ancilla is fixed and any input state can be cloned, but imperfectly.   

 In this work, we propose a state-dependent quantum copying:
 
  \begin{equation}
  	|\Psi\rangle \otimes |A_\Psi\rangle\xrightarrow{\mathcal{U}}|\Psi\rangle \otimes |\Psi\rangle
  	\label{cloning}
  	\end{equation}
  where $\ket{A_{\Psi}} $ is an ancilla which implicitly depends on $\ket{\Psi}$. As we will see this  formulation does not violate the no-cloning theorem because unlike universal cloning, the ancilla $\ket{A_{\Psi}} $ is state-dependent and not fixed. Further, unlike universal cloning that fail due to inherent non-linearity or resort to approximate copying \cite{buvzek1996quantum}, our construction defines a linear and unitary map across the Hilbert space by extending its action from a chosen orthonormal basis.  In our framework, we shall treat $\ket{A_\Psi}$ as an \emph{adaptive ancilla}. We define it as \emph{an ancillary quantum system whose state depends implicitly on the quantum state $|\Psi\rangle$ to be cloned — not through prior engineering or classical encoding, but through dynamic alignment governed by the system’s physical interaction}\footnote{Throughout this work, “implicit information” refers exclusively to the symmetry-governed structural capacity of the ancilla to host a copy, and not to any classical preparation or pre-encoding of the amplitudes or parameters of the unknown quantum state.}. 
 
 \textit{Stimulated emission} is well known physical process in quantum optics\cite{grynberg2010introduction}.  In the earlier studies\cite{simon2000cloning}, it has been theoretically  shown that stimulated emission in a suitably symmetric medium can implement an optimal universal cloning machine, albeit an approximate one. However, these  studies addressed universal cloning with inevitable imperfections. In this paper, we reframe the stimulated emission process in the language of quantum information theory. We  provide a novel interpretation of \textit{stimulated emission} as a physical realization of state-dependent quantum copying. The excited atom acts as an ancilla pre-aligned with the polarization of an incoming photon. When properly matched, this interaction results in the emission of a second photon in the same polarization. This resembles cloning, but is allowed since the process is not universal.  An excited state of an atom is a physical realization of abstract adaptive ancilla $\ket{A_\Psi}$. It is not engineered to match the incoming state $\ket{\Psi}$; the physical interaction dynamically selects transitions where the internal structure of the atom aligns with $\ket{\Psi}$, enabling stimulated emission. Thus, the stimulated emission realizes the same transformation as the abstract scheme (\ref{cloning}), thereby suggesting an interpretation of the excited state as an \emph{adaptive ancilla}. This perspective highlights that state-dependent cloning can occur naturally without prior classical preparation of the ancilla. 

We shall also discuss  the limits of the state-dependent quantum copying via adaptive ancilla. We shall argue that the cloning in case of stimulated emission processes are physically enforced by symmetry constraints in the atom rather than the no-cloning theorem alone. These symmetries are the restrictions on the top of basic structure of quantum mechanics and not all physical systems are constrained by the same symmetries. Thus one may imagine symmetry-relaxed physical systems, for instance Rydberg atoms, which supports a larger set of states that can be cloned via state-dependent copying mechanism.

Thus our formulation is grounded in a physically realizable system, drawing direct inspiration from light-matter interaction dynamics, thereby providing a natural and operational framework where cloning-like behavior arises without violating the no-cloning theorem. 

We organize the paper as follow: In section \ref{Formal Structure} we give formal structure of our state-dependent cloning via adaptive-ancilla. In section \ref{Light-Matter Interaction}A we discuss our cloning proposal in the context of stimulated emission. In section \ref{Light-Matter Interaction}B we discuss the cloning fidelity and compare it with the optimal universal cloning scheme. In section \ref{Limits} we discuss the limits of adaptive-ancilla assisted state-dependent cloning from group theoretic perspective. Finally in section \ref{Summary} we summarize and conclude.

\section{Formal Structure of State-Dependent Copying}
\label{Formal Structure}

 Let $\mathcal{H}^{S}$ and $\mathcal{H}^{A}$ be the Hilbert spaces of the two subsystems. For simplicity, we consider a two dimensional Hilbert space $\mathcal{H}_2$. $\mathcal{H}^{S}$  is spanned by   $\{|\psi_1\rangle, \ket{\psi_2}\}$ and $\mathcal{H}^{A}$ is spanned either by $\{|\psi_1\rangle, \ket{\psi_2}\}$ or $\{|A_{\psi_1}\rangle, \ket{A_{\psi_2}}\}$ . Consider a  transformation $\mathcal{U}$ on the composite system ( i.e on the combined Hilbert space $\mathcal{H}^{S}\otimes\mathcal{H}^{A}$) such that,

\begin{equation}
	|\psi_1\rangle\otimes|A_{\psi_1}\rangle\xrightarrow{\mathcal{U}}|\psi_1\rangle\otimes|\psi_1\rangle
	\label{cl11}
\end{equation} 

\begin{equation}
	|\psi_2\rangle\otimes|A_{\psi_2}\rangle\xrightarrow{\mathcal{U}}|\psi_2\rangle\otimes|\psi_2\rangle
	\label{cl22}
\end{equation}

\begin{equation}
	|\psi_1\rangle\otimes|A_{\psi_2}\rangle\xrightarrow{\mathcal{U}}|\psi_1\rangle\otimes|\psi_2\rangle
	\label{cl12}
\end{equation}

\begin{equation}
	|\psi_2\rangle\otimes|A_{\psi_1}\rangle\xrightarrow{\mathcal{U}}|\psi_2\rangle\otimes|\psi_1\rangle.
	\label{cl21}
\end{equation}

$\mathcal{U}$ is a unitary transformation as it maps orthonormal basis to another orthonormal basis. Let the general input state be:

\begin{equation}
|\Psi\rangle = \alpha |\psi_1\rangle + \beta |\psi_2\rangle, \quad |A_\Psi\rangle = \alpha |A_{\psi_1}\rangle + \beta |A_{\psi_2}\rangle
\end{equation}

Note that the we have same expansion coefficients $\alpha, \beta$  in $\ket{\Psi}$ as well as $\ket{A_{\Psi}}$. However, it does not imply that the ancilla explicitly contains the information about the values of $\alpha$ and $\beta$, it merely reflects linearity of a map: $V : \mathcal{H}^S_n \rightarrow \mathcal{H}^A, \quad V|\psi_i\rangle = |A_{\psi_i}\rangle$. Then by linearity of quantum mechanics:
\begin{equation}
|\Psi\rangle \otimes |A_\Psi\rangle \xrightarrow{\mathcal{U}} |\Psi\rangle \otimes |\Psi\rangle
\end{equation}

This is the state dependent copying via adaptive ancillary system. Note that, at formal mathematical level, the map $\ket{\Psi}\mapsto\ket{A_{\Psi}} $  is linear and exhaust the full Hilbert space.  Further, it is straightforward  generalize this map to n-dimensional Hilbert space (see appendix \ref{app1}). However, in the physical implementation, only symmetry restricted proper subspace $\mathcal{D}_{\text{clone}}$ of the full Hilbert space $\mathcal{H}^{S}$ participate in the cloning process. We shall  discuss this in the next section in the context of stimulated emission which we re-interpret as a physical system in which adaptive-ancilla assisted state-dependent cloning takes place.   We stress that our construction does not conflict with the no-cloning theorem: the theorem forbids a single universal machine that clones arbitrary unknown states. It does not forbid cases where a copy or the information necessary for copying pre-exists in implicit form within the ancilla. Our adaptive ancilla framework precisely exploits this possibility. Our construction thus establishes a rigorous and mathematically consistent way to implement quantum copying over any finite-dimensional Hilbert space using a state-aligned ancilla.

\section{Light-Matter Interaction as a State-Dependent Copying Process}
\label{Light-Matter Interaction}
\subsection{Stimulated Emission as Conditional State-Dependent Copying}

In quantum optics \cite{scully1997quantum,grynberg2010introduction,gerry2023introductory}, stimulated emission occurs when an excited atom interacts with an incoming photon and emits a second photon with identical frequency, direction, and polarization. The interaction Hamiltonian is\cite{scully1997quantum}:
\begin{equation}
\hat{H}_{\text{int}} = -\hat{\mathbf{p}} \cdot \hat{\mathbf{E}}=-\sum_\gamma \hat{\mathbf{p}} \cdot \boldsymbol{\epsilon}_\gamma  (\hat{a}_\gamma+\hat{a}^{\dagger}_{\gamma})
\end{equation}
where $\hat{a}^{\dagger}_\gamma$ and $\hat{a}_\gamma$ are the creation and annihilation operators, \( \hat{\mathbf{p}} \) is the atomic dipole moment, and \( \hat{\mathbf{E}} \) is the quantized electric field. The transition amplitude is:
\begin{equation}
\mathcal{M}_{ge} \propto \langle g | \hat{\mathbf{p}} \cdot \boldsymbol{\epsilon}_\gamma | e \rangle
\end{equation}
where $\bf{\epsilon}_\gamma$ is the polarization vector of the incoming photon state $\ket{\gamma}$.  $\ket{g}$ and $\ket{e}$ are the ground and excited state of an atom respectively. Efficient emission occurs only when the incoming photon's polarization state \( \ket{\gamma} \) aligns with the allowed dipole transition.  To clarify the adaptive nature of the ancilla, it is important to emphasize that the atomic system does not contain any pre-encoded information about the incoming photon polarization. The
dependence on the photon state arises entirely from the structure of the dipole interaction itself. To see this, consider an incoming photon polarization in the state $\ket{\gamma}$ which can be expressed as a linear combination
\begin{equation}
|\gamma\rangle = \sum_{i} \alpha_i |\lambda_i\rangle,
\end{equation}
where $\alpha_i $ are the unknown complex amplitudes and $|\lambda_i\rangle$ denote polarization basis states (e.g., circular or linear polarizations) which spans a symmetry-allowed subspace of the full Hilbert space (this subspace corresponds to $\mathcal{D}_{\text{clone}}$ as discussed in the next section).  
The dipole interaction Hamiltonian  couples  to polarization components $\bf{\epsilon_\lambda}$ for which the transition matrix element
\begin{equation}
\mathcal{M}_{ge}^{(\lambda_i)} \propto \langle g | \hat{\mathbf{p}}\cdot \boldsymbol{\epsilon}_{\lambda_i} | e \rangle
\end{equation}

is non-vanishing. All the polarization components orthogonal to the allowed dipole moment do not participate in the cloning dynamics as their transition amplitudes vanish identically. Thus any arbitrary polarization state $\ket{\gamma}$, the transition amplitude $\mathcal{M}_{ge}(\gamma)\propto \sum_{\lambda}\alpha_i\mathcal{M}_{ge}^{(\lambda_i)} $ and it is also faithfully cloned as long as it  belongs to $\mathcal{D}_{\text{clone}}$. This does not mean that the ancilla already contain {\it{explicit}} information about $\alpha_i$ or the state was pre-selected. The ancilla contain an {\it{implicit}} information in the form of a structured set of dynamical response channel constrained by the symmetries, one of which is activated by the unknown state via dynamical interaction. The photon polarization in unknown to the experimenter and the Hamiltonian dynamically selects the  channel compatible with the incoming photon quantum state.

This mechanism should be distinguished from state-dependent cloning schemes that rely on
pre-engineered ancilla states. In the present case, the excited atomic state $|e\rangle$ is fixed and
independent of the incoming photon polarization. The apparent state dependence arises solely
because stimulated emission can occur only along symmetry-allowed transition channels.
Consequently, the ancilla does not ``know'' the photon polarization in advance; rather, the
interaction Hamiltonian dynamically selects the compatible state from the symmetry-allowed subspace of polarization states. The excited atom state $\ket{e}$ thus acts as an adaptive ancilla that, when correctly aligned, allows:
\begin{equation}
|\gamma\rangle \otimes |e\rangle \xrightarrow{\mathcal{U}} |\gamma\rangle \otimes |\gamma\rangle\otimes\ket{g}
\end{equation}

This is the first known modeling of stimulated emission explicitly in terms of quantum information cloning dynamics reinterpreting an excited atom as an adaptive ancilla which holds a vast space of potential ancilla states, and selects the right one via interaction.

\subsection{Cloning Fidelity and Spontaneous Emission }
\label{Fidelity}
Although spontaneous emission can be interpreted as stimulated emission by the vacuum photons, it is not a cloning. The reason is that the vacuum contains all modes equally. The emitted photon's polarization is randomly distributed due to the lack of structure in the vacuum state. Although spontaneous emission arises from vacuum fluctuations, it does not yield a deterministic polarization state \( |\gamma\rangle \). Therefore, the excited atom plus vacuum field state \( |e\rangle \otimes |0\rangle \) cannot be equated with \( |\gamma\rangle \otimes |A_\gamma\rangle \). True copying requires a real incoming photon in state \( |\gamma\rangle \), not virtual quantum fluctuations.  Spontaneous emission is a background process that merely puts a bound on the fidelity of the cloning schemes \cite{herbert1982flash, milonni1982photons}.

In this context, it is interesting to contrast adaptive-ancilla assisted state-dependent cloning with that of optimal universal cloning\cite{buvzek1996quantum,simon2000cloning}. In the optical quantum cloning scheme, stimulated emission is employed to approximate a universal cloning machine that operates identically for all input polarization states\cite{simon2000cloning}. The device is treated as an amplifier-like system with fixed output modes. The unconditional (ensemble-averaged) single-clone fidelity is defined as $F_{\text{uncond}}=\langle\gamma|\rho_{\text{uncond}}^{\text{out}}|\gamma\rangle$, where $\ket{\gamma}$ is the state of the original photon qubit which lies within full Hilbert space and $\rho_{\text{uncond}}^{\text{out}}$ is the (unconditional) reduced density matrix  of one of the clones obtained by tracing over all output photons without postselection, i.e including all processes that contribute to the amplifier output. In this setting, spontaneous emission is inseparable from stimulated emission because both populate the same spatio-temporal output channel while differing only in polarization rendering $\rho_{\text{uncond}}^{\text{out}}$ generally mixed. As a result, photons emitted spontaneously—whose polarization is uncorrelated with the input—inevitably contribute "wrong-polarization” noise to the output ensemble. This noise lowers the $F_{\text{uncond}}$ below unity and enforces the optimal universal cloning bound. Thus in the universal optimal cloning scheme the  spontaneous emission fundamentally limit the cloning fidelity.

In contrast, in the adaptive-ancilla scheme cloning is defined as a conditional event: when the input polarization lies within the clonable domain $\mathcal{D}_{\text{clone}}$, stimulated emission produces a second photon in the same spatio-temporal and polarization mode as the input. The fidelity is defined as $F_{\text{cond}}=\langle\gamma|\rho_{\text{cond}}^{\text{out}}|\gamma\rangle$ where where $\ket{\gamma}\in D_{\text{clone}}$ and $\rho_{\text{cond}}^{\text{out}}$ is the (conditional) reduced density matrix which can be pure. This process yield unit fidelity conditioned on successful copying of the signal photon state. Spontaneous emission does not generate an incorrectly polarized clone in the target mode; instead, it corresponds to a failure event in which no clone is produced. Consequently, spontaneous emission limits only the success probability, not the conditional fidelity. This distinction—unconditional ensemble fidelity in optimal cloning scheme versus conditional, event-based fidelity in our scheme—fully explains why spontaneous emission reduces fidelity in universal optical cloning but does not contradict the unit-fidelity copying in our state-dependent framework.

\section{Physical limits of the state-dependent cloning via adaptive ancilla}
\label{Limits}
While the no-cloning theorem prohibits universal state duplication using a fixed ancilla (universal cloning), it does not preclude cloning-like behavior when the ancilla is constructed to match the state (state-dependent cloning). This work reveals a new class of processes—unitary, physically realizable, and state-dependent—which perform perfect copying without violating fundamental constraints of quantum mechanics. In the previous section we saw that the stimulated emission process can be interpreted  as a state-dependent cloner  because  an excited atom is an {\it{adaptive ancilla}} which holds a vast space of potential ancilla states, and selects the right one via interaction. However, this process adheres to the no-cloning theorem because  ancilla is not fixed but state-dependent. 

As discussed in the previous section, an excited atom cannot clone all the polarization states. In case of stimulated emission, the state-dependent cloning  is constrained by the selection rules which forbid certain transitions and hence not all the polarization states of the incoming  photon  are cloned through light-matter interactions\cite{grynberg2010introduction}. However, these restrictions are not encoded by the no-cloning theorem but the symmetries imposed on the atomic system, namely rotational symmetry and parity, etc thereby restricting the space of states that could be cloned.. These symmetries are imposed on the basic structure of quantum mechanics and they are not necessarily govern all the physical systems.  

\subsection{Group Theoretic Perspective of selection rules}

The selection rules are often derived from group representation theory. Let \( G \) be the symmetry group of the atomic system, and let \( \mathcal{H}^S \) and \( \mathcal{H}^A \) denote the Hilbert spaces of the system and ancilla respectively. Suppose \( \rho: G \to \mathrm{End}(\mathcal{H}^A) \) is a unitary representation of \( G \) acting on the ancilla states. Then, a dipole-allowed transition from an excited state \( |e\rangle \) to a ground state \( |g\rangle \) via an incoming photon in polarization state \( |\gamma\rangle \) is only permitted if the transition matrix element $\mathcal{M}_{ge}$ is non-zero. This condition is satisfied only when the irreducible representation (IR) \( \Gamma_g \) of \( |g\rangle \) appears in the tensor product \( \Gamma_e \otimes \Gamma_\gamma \), i.e.,

\begin{equation}
\Gamma_g \subset \Gamma_e \otimes \Gamma_\gamma,
\end{equation}
where \( \Gamma_e \) and \( \Gamma_\gamma \) are the IRs corresponding to the excited state and the photon's polarization, respectively. Therefore, only those input states \( |\gamma\rangle \) for which this condition is met can participate in cloning-like dynamics through stimulated emission. This defines the effective domain $\mathcal{D}_{\text{clone}}$ of the cloning transformation:

\begin{equation}
\mathcal{D}_{\text{clone}} = \left\{ |\gamma\rangle \in \mathcal{H}^S \, \middle| \, \mathcal{M}_{ge}\neq 0 \right\}.
\end{equation}

This domain is generally a proper subset of the full Hilbert space \( \mathcal{H}^S \). Hence, the limitation on the set of clonable quantum states arises not from the no-cloning theorem itself but from the symmetry constraints embedded in the structure of the physical ancilla system. This insight reveals that the true physical limits on cloning are imposed not by the impossibility of universal cloning per se, but by the dynamical laws and symmetry rules governing the system's evolution. 

In principle, if one could engineer a physical system where such symmetry restrictions are relaxed or bypassed, the set of states that can be effectively cloned using adaptive ancilla  could be significantly expanded. If one could design a physical system (e.g., exotic atoms, engineered qubits) where the group $G$ is trivial or its representations are sufficiently rich, then the map $\ket{\Psi}\otimes\ket{A_{\Psi}}\rightarrow\ket{\Psi}\otimes\ket{\Psi}$ could be realized over a much larger subspace.

\subsection{Rydberg Atom  as Realizations of Perfect State-Dependent Cloning}

Consider a Rydberg atom with a manifold of excited states \( \{ |e_i\rangle \} \) which constitute a larger subspace $\mathcal{R}_{\text{clone}}$ of $\mathcal{H}^s$, i.e $
	\mathcal{D}_{\text{clone}} \subset \mathcal{R}_{\text{clone}} \subset \mathcal H^S
$. Each $\ket{e_i}$ is dipole-coupled to a common ground state \( |g\rangle \). An incoming photon polarization state is \( |\gamma\rangle = \sum_i \alpha_i |\lambda_i\rangle \), where each \( |\lambda_i \rangle \) corresponds to a polarization mode that can couple to a specific transition \( |e_i\rangle \rightarrow |g\rangle \).

We associate the ancilla state $\ket{\mathcal{E}}$ with the excited atomic superposition:

\begin{equation}
|\mathcal{E}\rangle = \sum_i \alpha_i |e_i\rangle.
\end{equation}

where $\alpha_i$ are the unknown complex coefficients. Again, this ancilla is not engineered by an observer, rather it a vast space of of states $\{\ket{e_j}\}$ that can dynamically align with the photon polarization $\ket{\lambda_i}$. The linearly implies that any arbitrary state $\ket{\mathcal E}\in \mathcal{R}_{\text{clone}}$ is mapped with $\ket{\gamma}$  and hence can be cloned.  The amplitude for stimulated emission is:

\begin{equation}
\mathcal{M}_{g\mathcal E} \propto \langle g | \hat{\mathbf{p}} \cdot \vec{\epsilon}_\gamma |\mathcal{E}\rangle.
\end{equation}

\begin{figure}[h]
	\vspace{-0.4cm}
	\begin{center}
		\includegraphics[width=10cm,height=4cm]{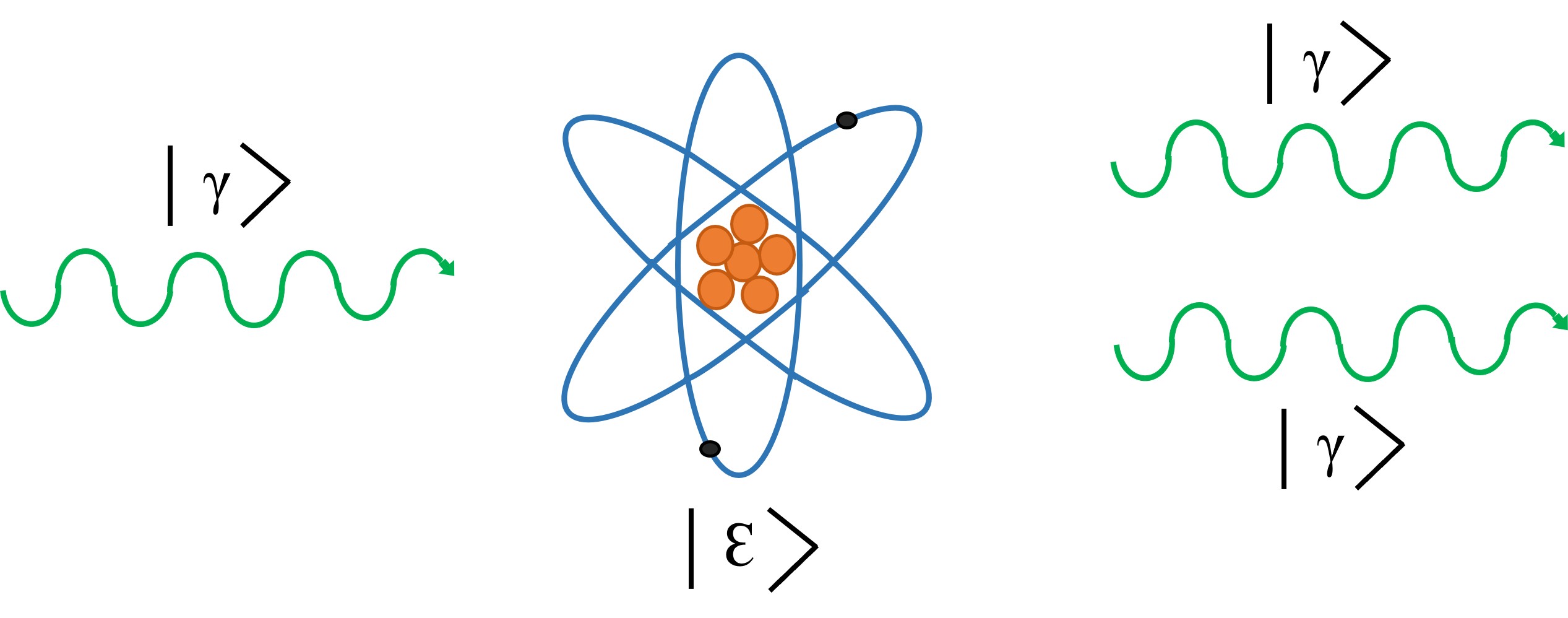}
		\caption{Figure illustrating the state-dependent quantum copying of a photon polarization  state $\ket{\gamma}$ via an excited atomic state $\ket{\mathcal E}$ (adaptive ancillary system). } 
		\label{stimulated1}
	\end{center}
\end{figure}

 Because all transitions \( |\psi_i\rangle \leftrightarrow |e_i\rangle \) are simultaneously supported with the non-zero transition amplitude $\mathcal{M}^{(\lambda_i)}_{g e_i}$, and the interaction is linear, the process:

\begin{equation}
|\gamma\rangle \otimes |\mathcal{E}\rangle \xrightarrow{\mathcal{U}} |\gamma\rangle \otimes |\gamma\rangle\otimes\ket{g}
\end{equation}
is effectively realized via unitary dynamics (see figure \ref{stimulated1} for the illustration of this process ). The ancilla state \( |\mathcal{E}\rangle \) is thus  selected by the interaction itself, not externally engineered. Furthermore, Rydberg atoms allow external tuning (e.g., via electric fields or microwave dressing) to break or relax symmetry-induced selection rules\cite{zhang2018symmetry} (e.g., \( \Delta j = 0, \pm1 \), \( \Delta m_j = 0, \pm1 \)). This enlarges the effective clonable domain \( \mathcal{D}_{\text{clone}} \subset \mathcal{H}^S \),  covering a larger subspace of the full Hilbert space.

Thus, Rydberg atoms provide a promising platform for exploring extended forms of state-dependent cloning via adaptive ancilla. Their large dipole moments and tunable level structures allow partial relaxation or engineering of selection rules through external fields or microwave dressing. While a detailed analysis is beyond the scope of the present work, such systems may enlarge the symmetry-allowed space $\mathcal{D}_{\text{clone}}$ and thereby reveal richer varieties of adaptive copying mechanisms. We frame this as a conceptual direction for future study.

\section{Conclusion and summary}
\label{Summary}
We introduced a novel state-dependent quantum cloning (copying) process by introducing a new class of ancillary system---an {\it{adaptive ancilla}}. This state-dependent ancillary system is not pre-engineered to match the quantum state to be cloned but contain an {\it{implicit}} structural information about it in the form of symmetry-restricted structured set of quantum states.  The cloning occurs when the atomic dipole state dynamically aligns with the quantum state to be cloned via interaction. We demonstrated that stimulated emission, a standard optical process, can be reinterpreted as a physically realizable model for state-dependent quantum copying via adaptive ancilla. The success of the process relies on matching the excited atom--an adaptive-ancilla of our scheme--to the input state (photon polarization). This does not violate the no-cloning theorem because the ancilla is not fixed.  If we reinterpret the atomic excited state as an adaptive ancilla then the stimulated emission process demonstrate the state-dependent  cloning  because  an excited atom  holds a vast space of potential ancilla states, and selects the right one via interaction. Although theoretically every photon polarization state can potentially be cloned in this formulation, symmetry governed forbidden transition put limits on the set of states that can be cloned using a given atomic system. We have to choose a different atomic system with many degrees of freedom so that a wide range of polarization states can be matched and hence cloned. This perspective may inspire new approaches to quantum amplification or state transfer that deliberately utilize state-aligned ancilla.

In conclusion, our formulation extends the foundational understanding of no-cloning theorem.  The universal cloning of an arbitrary quantum state is still  forbidden, however conditional copying via adaptive ancilla  is allowed.  Ancilla does not contain this information {\it{explicitly}} but only {\it{implicitly}} in the form of structured set of symmetry-allowed dynamical response channels. The framework developed in this paper  provides a unifying conceptual and practical foundation for next-generation quantum technologies that exploit dynamical alignment rather than universal programming.

\section*{Acknowledgment}

I especially like to thank Prof. Nicolas Gisin for encouraging comments on the basic premise of this work.  I would also like to acknowledge useful comments by  Prof. Richard Josza, Prof. Samuel Braunstein., Prof. Anirban Pathak, Dr Sumiran Pujari and Dr Sandeep Mishra. I also thank Swapnali Pawar for creating images.

\section*{Conflict of interests statement }
The authors declares that there is no conflict of interests regarding the publication of this manuscript.

\section*{Funding}
The authors declare that no funds, grants, or other support were received during the preparation of
this manuscript.

\section*{Data Availability}
No datasets were generated or analyzed in this study.

\appendix

\section{Linearity of the Mapping \( |\Psi\rangle \mapsto |A_\Psi\rangle \) in \( n \)-Dimensional Hilbert Space}
\label{app1}
Let \( \mathcal{H}^S_n \) be an \( n \)-dimensional Hilbert space with orthonormal basis \( \{ |\psi_i\rangle \}_{i=1}^n \). Define a unitary operator \( \mathcal{U} \) that acts on basis product states as follows:
\begin{equation}
|\psi_i\rangle \otimes |A_{\psi_j}\rangle \xrightarrow{\mathcal{U}} |\psi_i\rangle \otimes |\psi_j\rangle \quad \text{for all } i, j \in \{1, \dots, n\}.
\end{equation}

Let an arbitrary state in \( \mathcal{H}_n \) be written as:
\begin{equation}
|\Psi\rangle = \sum_{i=1}^n \alpha_i |\psi_i\rangle,
\quad
|A_\Psi\rangle = \sum_{j=1}^n \alpha_j |A_{\psi_j}\rangle.
\end{equation}

Then the total input state is:
\begin{equation}
|\Psi\rangle \otimes |A_\Psi\rangle = \sum_{i=1}^n \sum_{j=1}^n \alpha_i \alpha_j |\psi_i\rangle \otimes |A_{\psi_j}\rangle.
\end{equation}

Using the defined action of \(\mathcal{U}  \), we have:
\begin{equation}
 |\Psi\rangle \otimes |A_\Psi\rangle
= \sum_{i=1}^n \sum_{j=1}^n \alpha_i \alpha_j (|\psi_i\rangle \otimes |A_{\psi_j}\rangle)
\xrightarrow{\mathcal{U}} \sum_{i=1}^n \sum_{j=1}^n \alpha_i \alpha_j |\psi_i\rangle \otimes |\psi_j\rangle.
\end{equation}

This is simply the tensor product:
\begin{equation}
|\Psi\rangle \otimes |\Psi\rangle = \left( \sum_{i=1}^n \alpha_i |\psi_i\rangle \right) \otimes \left( \sum_{j=1}^n \alpha_j |\psi_j\rangle \right)
= \sum_{i,j} \alpha_i \alpha_j |\psi_i\rangle \otimes |\psi_j\rangle.
\end{equation}

Therefore, the overall transformation satisfies:
\begin{equation}
|\Psi\rangle \otimes |A_\Psi\rangle \xrightarrow{\mathcal{U}} |\Psi\rangle \otimes |\Psi\rangle.
\end{equation}

Define the ancilla preparation map:
\begin{equation}
V : \mathcal{H}^S_n \rightarrow \mathcal{H}^A, \quad V|\psi_i\rangle = |A_{\psi_i}\rangle.
\end{equation}
where $\mathcal{H}^S_n$ and $\mathcal{H}^A$ are the system and ancilla Hilbert spaces respectively. Then, for all \( |\Psi\rangle \in \mathcal{H}^S_n \),
\begin{equation}
|A_\Psi\rangle = V|\Psi\rangle,
\end{equation}
showing that the mapping \( |\Psi\rangle \mapsto |A_\Psi\rangle \) is linear. Note however that this linearity holds at the formal mathematical level, assuming the ancilla basis \( \{ |A_{\psi_i}\rangle \} \) is predefined. In physical implementation, engineering \( |A_\Psi\rangle \) for arbitrary unknown \( |\Psi\rangle \) would require  prior and exact knowledge of the state, which is restricted by the no-cloning theorem. However, in the adaptive-ancilla assisted state-dependent cloning scheme, it is possible to copy an arbitrary state as long as $\ket{\Psi}\in \mathcal{D}_{\text{clone}}$. In this scheme, as discussed in the paper, the ancilliary system $\ket{A_{\Psi}}$ does not contain explicit information about $\ket{\Psi}$ but only the implicit information in the form of structured set of dynamical response channels constrained by the symmetries.

\section{Group theoretic derivation of the selection rules }

\label{sec:selection_rules_group}

Let $G$ denote the symmetry group of the atom-field system. For an isolated atom in free space,
$G$ is generated by spatial rotations $SO(3)$ together with parity $\mathbb{Z}_2$. Atomic
eigenstates transform under irreducible representations (IRs) of $SO(3)$ labeled by angular
momentum quantum numbers $j$\cite{weyl1950theory}. $\Gamma_g=\Gamma^{(j_g)}$ is the IR of the atomic ground state and $\Gamma_e=\Gamma^{(j_e)}$ is the IR of an excited state   The electric dipole operator $\hat{\mathbf{p}}$ transforms as a vector operator (tensor of rank-1). Under a rotation $R\in SO(3)$, $\hat{\mathbf{p}}$  transforms as\cite{georgi2000lie}
\begin{equation}
	\hat U(R)\,\hat p_i\,\hat U^\dagger(R)= \sum_{k} R_{ik}\,\hat p_k ,
\end{equation}
showing that its components mix among themselves as a vector. Thus, $\hat{\mathbf{p}}$ carries the rank-1 irreducible representation of $SO(3)$.

It is convenient to express the dipole operator in the spherical tensor basis
$\{T^{(1)}_q\}_{q=-1}^{+1}$, where
\begin{equation}
	\hat{\mathbf{p}} \equiv \left\{ T^{(1)}_{-1},\, T^{(1)}_{0},\, T^{(1)}_{+1} \right\}.
\end{equation}
Under rotations a spherical tensor of rank k transform as\cite{georgi2000lie},

\begin{equation}
	\hat U(R)\, T^{(k)}_q \, \hat U^\dagger(R)
	=
	\sum_{q'=-k}^{+k} D^{(k)}_{q'q}(R)\, T^{(1)}_{q'},
\end{equation}

For $T^{(1)}_q$ we get

\begin{equation}
	\hat U(R)\, T^{(1)}_q \, \hat U^\dagger(R)
	=
	\sum_{q'=-1}^{+1} D^{(1)}_{q'q}(R)\, T^{(1)}_{q'},
\end{equation}
where $D^{(1)}_{q'q}(R)$ are the Wigner $D$-matrices. Thus the dipole operator transform as $\Gamma^{(1)}$.

The transition amplitude for absorption or emission of a photon with polarization state
$|\gamma\rangle$ is
\begin{equation}
	M_{ge}(\gamma)
	\propto
	\langle g | \hat{\mathbf{p}}\cdot \boldsymbol{\epsilon}_\gamma | e \rangle
	=
	\sum_{q=-1}^{+1}
	\epsilon_{\gamma,-q}\,
	\langle g | T^{(1)}_q | e \rangle =\sum_{q=-1}^{+1}
	\epsilon_{\gamma,-q}\,
	\langle j_g m_g | T^{(1)}_q | j_e m_e \rangle,
\end{equation}
where $\boldsymbol{\epsilon}_\gamma$ is the polarization vector (tensor of rank 1 i.e it also transform as $\Gamma^{(1)}$) and
$\epsilon_{\gamma,q}$ its spherical components. $m$ is the magnetic quantum  number.

Note that the angular momentum conservation couples the photon polarization state with that of electric dipole moment as
\begin{equation}
	T^{(1)}_{q} \;\longleftrightarrow\; \epsilon_{\gamma,-q},
	\quad
	q=+1\,(\text{LCP}),\;
	q=0\,(\pi),\;
	q=-1\,(\text{RCP}).
\end{equation}
where $\pi$ is the linear polarization state of the incoming photon and $\text{LCP, RCP}$ corresponds to left and right circular polarization respectively.

 \begin{figure}[h]
	\vspace{-0.4cm}
	\begin{center}
		\includegraphics[width=10cm,height=4cm]{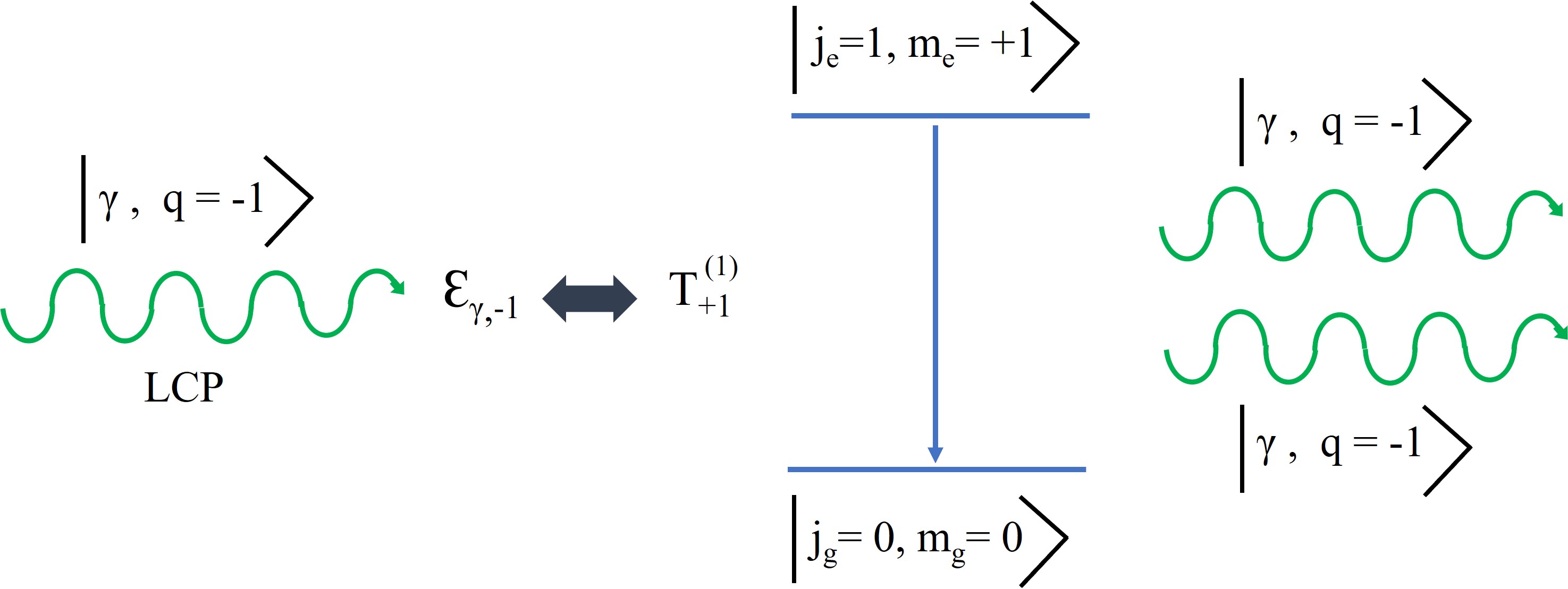}
		\caption{Figure illustrating the state-dependent quantum copying of a left circularly polarized (LCP) photon  state $\ket{\gamma,q=-1}$ via an excited atomic state $\ket{j_e=1,m_e=+1}$ (adaptive ancillary system).  } 
		\label{stimulated}
	\end{center}
\end{figure}

By the Wigner--Eckart theorem\cite{georgi2000lie},
\begin{equation}
	\langle j_g m_g | T^{(1)}_q | j_e m_e \rangle
	=
	\langle j_g \| T^{(1)} \| j_e \rangle
	\, C^{j_g m_g}_{j_e m_e,\, 1 q},
\end{equation}
where $C^{j_g m_g}_{j_e m_e,\, 1 q}$ are Clebsch--Gordan coefficients. Consequently, the matrix
element is nonzero only when the corresponding Clebsch--Gordan coefficient is nonvanishing,
yielding the familiar dipole selection rules
\begin{equation}
	\Delta j = 0,\pm 1 \quad (j_e = j_g = 0 \ \text{forbidden}), 
	\qquad
	\Delta m = q = 0,\pm 1 .
	\label{selection}
\end{equation}
In addition, electric dipole transitions require a change of parity,
\begin{equation}
	\pi_g = -\,\pi_e .
\end{equation}

Selection rules (\ref{selection}) can also be deduced from the fact that a necessary and sufficient condition for a non vanishing dipole matrix element $\mathcal{M}_{ge}$ is
\begin{equation}
	\Gamma_g^\ast \otimes \Gamma_{\mathrm{dip}} \otimes \Gamma_e
	\;\supset\;
	\Gamma_{\mathrm{triv}},
\end{equation}
where $\Gamma_{\text{dip}}$ is the IR of dipole moment operator  and $\Gamma_{\text{triv}}$ is the trivial representation. Equivalently,
\begin{equation}
	\Gamma_g \subset \Gamma_{\mathrm{dip}} \otimes \Gamma_e .
\end{equation}
For $SO(3)$ this reduces to the angular momentum addition rule
\begin{equation}
	j_e \otimes 1 = (j_e-1) \oplus j_e \oplus (j_e+1),
\end{equation}
which  finally give the selection rules $\Delta j=0,\pm1$.
\bibliography{cloning}

\end{document}